\documentclass{amsart}%
\usepackage{amsfonts}
\usepackage{amsmath}
\usepackage{amssymb}
\usepackage{graphicx}%
\providecommand{\U}[1]{\protect\rule{.1in}{.1in}}
\theoremstyle{plain}
\newtheorem{theorem}{Theorem}[section]

\newtheorem{proposition}[theorem]{Proposition}
\theoremstyle{definition}

\theoremstyle{remark}
\newtheorem{remark}[theorem]{Remark}
\numberwithin{equation}{section}

\begin{document}
\title[On the complete integrability of a Riemann type hydrodynamic system]{On the complete Lax type integrability of a generalized Riemann type
hydrodynamic system}
\author{Denis Blackmore}
\address{Department of Mathematical Sciences at the New Jersey Institute of Technology
(NJIT), Newark NJ 07102, USA }
\email{deblac@m.njit.edu}
\author{Yarema A. Prykarpatsky}
\address{The Department of Applied Mathematics at the Agrarian University of Cracow,
Poland, the Institute of Mathematics at the NAS, Kyiv, Ukraine, the Department
Mathematics at the Ivan Franko Pedagogical State University, Drohobych, Lviv
region, Ukraine }
\email{yarpry@gmail.com}
\author{Orest D. Artemowych}
\address{The Department of Algebra and Topology at the Faculty of Mathematics and
Informatics of the Vasyl Stefanyk Pre-Carpathian National University,
Ivano-Frankivsk, Ukraine, and the Institute of Mathematics and Informatics at
the Tadeusz Kosciuszko University of Technology, Cracow, Poland}
\email{artemo@usk.pk.edu.pl}
\author{Anatoliy K. Prykarpatsky}
\address{AGH University of Science and Technology, Krakow \\
30059 Poland, and the Ivan Franko Pedagogical State University, Drohobych,
Lviv region, Ukraine }
\email{pryk.anat@ua.fm, prykanat@cybergal}
\date{present}

\begin{abstract}
The complete integrability of a generalized Riemann type hydrodynamic system
is studied by means of a novel combination of symplectic and
differential-algebraic tools. A compatible pair of polynomial Poissonian
structures, a Lax representation and a related infinite hierarchy of
conservation laws are constructed.

\end{abstract}
\maketitle

\section{Introduction}

We shall study the complete integrability of the dispersionless Riemann type
hydrodynamic flow%
\begin{equation}
D_{t}^{N-1}u=\bar{z}_{x}^{2},\text{ \ \ \ }D_{t}\bar{z}=0 \label{N1.1}%
\end{equation}
on a $2\pi$-periodic functional manifold $\bar{M}^{N}\subset C^{(\infty
)}(\mathbb{R}/2\pi\mathbb{Z};\mathbb{R}^{N}),$ where $N\in\mathbb{N}$ is an
arbitrary\ natural number, the vector $(u,D_{t}u,D_{t}^{2}u,...,D_{t}%
^{N-1}u,\bar{z})^{\intercal}\in\bar{M}^{N},$ the differentiations
$D_{x}:=\partial/\partial x,$ $D_{t}:=\partial/\partial t+u\partial/\partial
x$ satisfy the Lie-algebraic commutator relationship%

\begin{equation}
\lbrack D_{x},D_{t}]=u_{x}D_{x}.\label{N1.1a}%
\end{equation}
and $t\in\mathbb{R}$ is an evolution parameter. \ The system can be considered
as a slight generalization of the dispersionless Riemann hydrodynamic system
(suggested recently by M. Pavlov and D. Holm \cite{PH}) in the form
\begin{equation}
D_{t}^{N-1}u=\bar{z},\text{ \ \ \ }D_{t}\bar{z}=0\label{N1.2}%
\end{equation}
for $N\in\mathbb{N}$ \ and extensively studied in
\cite{PAPP,Po,GBPPP,GPP,PoP,PAPP-1}, \ where it was proved that it is a Lax
integrable bi-Hamiltonian flow on the manifold $\bar{M}^{N}$ and possesses an
infinite hierarchy of mutually commuting \ dispersive Lax integrable
Hamiltonian flows.

For the case $N=2$ it is well known \cite{PM,BPS} that the system (\ref{N1.1})
is a smooth Lax integrable bi-Hamiltonian flow on the $2\pi$-periodic
functional manifold $\bar{M}^{2},$ \ whose Lax representation is given by the
compatible linear system%
\begin{equation}
D_{x}f=\left(
\begin{array}
[c]{cc}%
\bar{z}_{x} & 0\\
-\lambda(u+u_{x}/\bar{z}_{x}) & -\bar{z}_{xx}/\bar{z}_{x}%
\end{array}
\right)  f,\text{ \ \ \ \ }D_{t}f=\left(
\begin{array}
[c]{cc}%
0 & 0\\
-\lambda\bar{z}_{x} & u_{x})
\end{array}
\right)  f, \label{N1.2a}%
\end{equation}
where $f\in C^{(\infty)}(\mathbb{R}^{2};\mathbb{R}^{2})$ $\ $and
$\ \ \ \lambda\in\mathbb{R}\ $is an arbitrary spectral parameter.

Our focus here is an investigation of the Lax integrability of the Riemann
type hydrodynamic system (\ref{N1.1}) for $N=3$ on a $2\pi$-periodic
functional manifold $\bar{M}^{3}\subset C^{(\infty)}(\mathbb{R}/2\pi
\mathbb{Z};\mathbb{R}^{3})$ for a vector $(u,v,\bar{z})^{\intercal}\in\bar
{M}^{3}$. We treat this problem in the following extended form:
\begin{equation}
D_{t}u=v,\text{ \ \ \ \ \ }D_{t}v=\bar{z}_{x}^{2},\text{ \ \ \ }D_{t}\bar
{z}=0. \label{N1.3}%
\end{equation}
The flow (\ref{N1.3}) can be recast as a one on a $2\pi$-periodic functional
manifold $M^{3}\subset C^{(\infty)}(\mathbb{R}/2\pi\mathbb{Z};\mathbb{R}^{3})$
for a vector $(u,v,z)^{\intercal}\in M^{3}$ as
\begin{equation}
D_{t}u=v,\text{ \ \ \ \ \ \ }D_{t}v=z,\text{ \ \ \ \ \ }D_{t}z=-2zu_{x},
\label{N1.3a}%
\end{equation}
where, for further convenience, we have made the change of variables:
$z:=\bar{z}_{x}^{2}.$ \ \ We will also use the form of the flow \ (\ref{N1.3a}%
):
\begin{equation}
\left(
\begin{array}
[c]{c}%
du/dt\\
dv/dt\\
dz/dt
\end{array}
\right)  =K[u,v,z]:=\left(
\begin{array}
[c]{c}%
v-uu_{x}\\
z-uv_{x}\\
-2u_{x}z-uz_{x}%
\end{array}
\right)  \label{N1.3b}%
\end{equation}
defining a standard smooth dynamical system on the infinite-dimensional
functional manifold $M^{3},$ where $K:M^{3}\rightarrow T(M^{3})$ is the
corresponding smooth vector field on $M^{3}.$\ 

In the sequel, we shall prove the following result using symplectic
gradient-holonomic and differential algebraic tools.

\begin{proposition}
\label{Pr_N1.0}The Riemann type hydrodynamic flow (\ref{N1.3b}) is a
bi-Hamiltonian dynamical system on the functional manifold $M^{3}$ with
respect to two compatible Poissonian structures $\vartheta,\eta:T^{\ast}%
(M^{3})$ $\rightarrow T(M^{3})$
\begin{equation}
\vartheta:=\left(
\begin{array}
[c]{ccc}%
0 & 1 & 0\\
-1 & 0 & 0\\
0 & 0 & 2z^{1/2}D_{x}z^{1/2}%
\end{array}
\right)  ,\eta:=\left(
\begin{array}
[c]{ccc}%
\partial^{-1} & u_{x}\partial^{-1} & 0\\
\partial^{-1}u_{x} & v_{x}\partial^{-1}+\partial^{-1}v_{x} & \partial
^{-1}z_{x}-2z\\
0 & z_{x}\partial^{-1}+2z & 0
\end{array}
\right)  , \label{N1.3c}%
\end{equation}
possessing an infinite hierarchy of mutually commuting conservation laws and a
non-autonomous Lax representation of the form
\begin{align}
D_{t}f  &  =\left(
\begin{array}
[c]{ccc}%
0 & 0 & 0\\
-\lambda & 0 & 0\\
0 & -\lambda z_{x} & u_{x}%
\end{array}
\right)  f,\text{ }\label{N1.3d}\\
D_{x}f  &  =\left(
\begin{array}
[c]{ccc}%
\lambda^{2}u\sqrt{z} & \lambda v\sqrt{z} & z\\
-\lambda^{3}tu\sqrt{z} & -\lambda^{2}tv\sqrt{z} & -\lambda tz\\%
\begin{array}
[c]{c}%
\lambda^{4}(tuv-u^{2})-\\
-\lambda^{2}u_{x}/\sqrt{z}%
\end{array}
&
\begin{array}
[c]{c}%
-\lambda v_{x}/\sqrt{z}+\\
+\lambda^{3}(tv^{2}-uv)
\end{array}
&
\begin{array}
[c]{c}%
\lambda^{2}\sqrt{z}(u-tv)-\\
-z_{x}/2z
\end{array}
\end{array}
\right)  f,\nonumber
\end{align}
where $\lambda\in\mathbb{R}$ is an arbitrary spectral parameter and $f\in
C^{(\infty)}(\mathbb{R}^{2};\mathbb{R}^{3}).$
\end{proposition}

\bigskip

\section{Symplectic gradient-holonomic integrability analysis}

Our first steps in proving Proposition 1.1 are fashioned using the symplectic
gradient-holonomic method, which takes us a long way towards the desired result.

\subsection{Poissonian structure analysis on the functional manifold $M^{3}$}

By employing the symplectic gradient-holonomic approach \cite{PM,BPS,HPP} to
studying the integrability of smooth nonlinear dynamical systems on functional
manifolds, one can find a set of conservation laws for \ (\ref{N1.3b}) by
constructing some solutions $\varphi:=\varphi\lbrack u,v,z]\in T^{\ast}%
(M^{3})$ to the functional Lax gradient equation:%
\begin{equation}
d\varphi/dt+K^{\prime,\ast}\varphi=\mathrm{grad}\mathcal{L}, \label{N1.4}%
\end{equation}
where $\varphi^{\prime}=\varphi^{\prime,\ast},$ $\mathcal{L}\in D(M^{3})$ is a
suitable Lagrangian functional and the linear operator $K^{\prime,\ast
}:T^{\ast}(M^{3})\rightarrow T^{\ast}(M^{3})$ is the adjoint with respect to
the standard convolution $(\cdot,\cdot)$ on $T^{\ast}(M^{3})\times T(M^{3}),$
of the Fr\'{e}chet-derivative of a nonlinear mapping $K:M^{3}\rightarrow
T(M^{3})$; namely,%
\begin{equation}
K^{\prime,\ast}=\left(
\begin{array}
[c]{ccc}%
uD_{x} & -v_{x} & z_{x}+2zD_{x}\\
1 & u_{x}+uD_{x} & 0\\
0 & 1 & -u_{x}+uD_{x}%
\end{array}
\right)  . \label{N1.5}%
\end{equation}
\ \ The Lax gradient equation \ (\ref{N1.4}) can be, owing to \ (\ref{N1.2}),
rewritten as%
\begin{equation}
D_{t}\varphi+k[u,v,z]\varphi=\mathrm{grad}\text{ }\mathcal{L}, \label{N1.6}%
\end{equation}
where the matrix operator
\begin{equation}
k[u,v,z]:=\left(
\begin{array}
[c]{ccc}%
0 & -v_{x} & z_{x}+2zD_{x}\\
1 & u_{x} & 0\\
0 & 1 & -u_{x}%
\end{array}
\right)  . \label{N1.7}%
\end{equation}
The first vector elements
\begin{align}
\varphi_{\vartheta}[u,v,z]  &  =(z-uv_{x},-v+uu_{x},u),\mathcal{L}_{\vartheta
}=0\label{N1.8}\\
\varphi_{\eta}[u,v,z]  &  =(v_{x},-u_{x},-1)^{\intercal},\mathcal{L}_{\eta
}=0,\nonumber\\
\varphi_{0}[u,v,z]  &  =(-(u_{x}z^{-1/2})_{x},(z^{-1/2})_{x},(v_{x}%
/2-u_{x}^{2}/4)z^{-3/2})^{\intercal},\mathcal{L}_{0}=0,\nonumber
\end{align}
as can be easily checked, are solutions of the functional equation
\ (\ref{N1.6}). From an application of the standard Volterra homotopy formula%
\begin{equation}
H:=\int_{0}^{1}d\mu(\varphi\lbrack\mu u,\mu v,\mu z],(u,v,z)^{\intercal}),
\label{N1.9}%
\end{equation}
one finds the conservation laws for \ (\ref{N1.2}); namely,%
\begin{align}
H_{\eta}  &  =\frac{1}{2}\int_{0}^{2\pi}dx(2uz-v^{2}-u^{2}v_{x}),
\label{N1.10}\\
H_{\vartheta}  &  :=\int_{0}^{2\pi}dx(uv_{x}/2-vu_{x}/2-z),\text{ \ \ }%
H_{0}:=\frac{1}{2}\int_{0}^{2\pi}dx(u_{x}^{2}-2v_{x})z^{-1/2}.\nonumber
\end{align}

It is now quite easy, making use of the conservation laws \ (\ref{N1.10}), to
construct a Poissonian structure $\vartheta:T^{\ast}(M^{3})\rightarrow
T(M^{3})$ for the dynamical system\ (\ref{N1.3b}). If we use the
representations
\begin{align}
H_{\vartheta}  &  =\int_{0}^{2\pi}dx(uv_{x}/2-vu_{x}/2-z):=(\psi_{\vartheta
},(u_{x},v_{x},z_{x})^{\intercal}),\label{N1.11}\\
\psi_{\vartheta}  &  :=(-v/2,u/2,z^{-1/2}D_{x}^{-1}z^{1/2}/2)^{\intercal
},\nonumber
\end{align}
it follows that the vector $\psi_{\vartheta}\in T^{\ast}(M^{3})$ satisfies the
Lax gradient equation (\ref{N1.6}):%
\begin{equation}
D_{t}\psi_{\vartheta}+k[u,v,z]\psi_{\vartheta}=\mathrm{grad}\text{
}\mathcal{L}_{\vartheta}, \label{N1.11a}%
\end{equation}
where the Lagrangian function $\ \mathcal{L}_{\vartheta}=(\psi_{\vartheta
},K)-H_{\vartheta}.$ Thus, based on the inverse co-symplectic functional
expression
\begin{equation}
\vartheta^{-1}:=\psi_{\vartheta}^{\prime}-\psi_{\vartheta}^{\prime,\ast
}=\left(
\begin{array}
[c]{ccc}%
0 & -1 & 0\\
1 & 0 & 0\\
0 & 0 & z^{-1/2}D_{x}^{-1}z^{-1/2}/2
\end{array}
\right)  \label{N1.12}%
\end{equation}
one readily obtains the linear co-symplectic operator on the manifold $M^{3}:$%

\begin{equation}
\vartheta:=\left(
\begin{array}
[c]{ccc}%
0 & 1 & 0\\
-1 & 0 & 0\\
0 & 0 & 2z^{1/2}D_{x}z^{1/2}%
\end{array}
\right)  , \label{N1.13}%
\end{equation}
which is the corresponding Poissonian operator for the dynamical
system\ (\ref{N1.1a}). \ It is also important to observe that the dynamical
system \ (\ref{N1.1a}) is a Hamiltonian flow on the functional manifold
$M^{3}$ with respect to the Poissonian structure \ (\ref{N1.13}).%

\begin{equation}
K[u,v,z]=-\vartheta\text{ }\mathrm{grad}\text{ }H_{\eta}. \label{N1.13a}%
\end{equation}

\subsection{Poissonian structure analysis on $\bar{M}^{3}$}

In what follows, we shall find it convenient to construct we will construct
other Poissonian structures for dynamical system (\ref{N1.3}) on the manifold
$\ \bar{M}^{3},$ rewritten in the equivalent form%
\begin{equation}
\frac{d}{dt}\left(
\begin{array}
[c]{c}%
u\\
v\\
\bar{z}%
\end{array}
\right)  =\bar{K}[u,v,\bar{z}]:=\left(
\begin{array}
[c]{c}%
v-uu_{x}\\
\bar{z}_{x}^{2}-uv_{x}\\
0
\end{array}
\right)  , \label{N1.3bb}%
\end{equation}
where $\bar{K}:\bar{M}^{3}\rightarrow T(\bar{M}^{3})$ is the corresponding
vector field on $\ \bar{M}^{3}.$ To proceed, we need to obtain additional
solutions to the Lax gradient equation (\ref{N1.6}) on the functional manifold
$\bar{M}^{3}$
\begin{equation}
D_{t}\bar{\psi}+\bar{k}[u,v,z]\bar{\psi}=\mathrm{grad}\text{ }\mathcal{\bar
{L}}, \label{N1.6a}%
\end{equation}
where the matrix operator is
\begin{equation}
\bar{k}[u,v,\bar{z}]:=\left(
\begin{array}
[c]{ccc}%
0 & -v_{x} & -\bar{z}_{x}\\
1 & u_{x} & 0\\
0 & -2\partial\text{ }\bar{z}_{x} & u_{x}%
\end{array}
\right)  , \label{N1.6b}%
\end{equation}
and which we may rewrite in the componentwise form
\begin{align}
D_{t}\bar{\psi}^{(1)}  &  =v_{x}\bar{\psi}^{(2)}+\bar{z}_{x}\bar{\psi}%
^{(3)}+\delta\mathcal{\bar{L}}/\delta u,\label{N1.13aa}\\
D_{t}\bar{\psi}^{(2)}  &  =-\bar{\psi}^{(1)}-u_{x}\bar{\psi}^{(2)}%
+\delta\mathcal{\bar{L}}/\delta v,\nonumber\\
D_{t}\bar{\psi}^{(3)}  &  =2(\bar{z}_{x}\bar{\psi}^{(2)})_{x}-u_{x}\bar{\psi
}^{(3)}+\delta\mathcal{\bar{L}}/\delta\bar{z},\nonumber
\end{align}
where the vector $\bar{\psi}:=(\bar{\psi}^{(1)},\bar{\psi}^{(2)},\bar{\psi
}^{(3)})^{\intercal}\in T^{\ast}(\bar{M}^{3}).$ As a simple consequence of
(\ref{N1.13aa}), one obtains the following \ system of differential
relationships:%
\begin{equation}%
\begin{array}
[c]{c}%
D_{t}^{3}\tilde{\psi}^{(2)}=-2\bar{z}_{x}^{2}\tilde{\psi}_{x}^{(2)}+D_{t}%
^{2}\partial^{-1}(\delta\mathcal{\bar{L}}/\delta v)-\\
-\partial^{-1}<\mathrm{grad}\text{ }\mathcal{\bar{L}},(u_{x},v_{x},\bar{z}%
_{x})^{\intercal}>,\text{ \ }\\
D_{t}\tilde{\psi}^{(2)}=-\tilde{\psi}^{(1)}+\partial^{-1}(\delta
\mathcal{\bar{L}}/\delta v),\\
D_{t}\tilde{\psi}^{(3)}=2\bar{z}_{x}\tilde{\psi}_{x}^{(2)}+\partial
^{-1}(\delta\mathcal{\bar{L}}/\delta\bar{z}).
\end{array}
\label{N1.13ab}%
\end{equation}
Here we have defined $(\bar{\psi}^{(1)},\bar{\psi}^{(2)},\bar{\psi}%
^{(3)})^{\intercal}:=(\tilde{\psi}_{x}^{(1)},\tilde{\psi}_{x}^{(2)}%
,\tilde{\psi}_{x}^{(3)})^{\intercal}$ and made use of the commutator
relationship for differentiations $D_{t}$ and $D_{x}:$%
\begin{equation}
\lbrack D_{t},\alpha^{-1}D_{x}]=0, \label{N1.13ba}%
\end{equation}
which holds for the function $\alpha:=1/\bar{z}_{x},$ where $D_{t}\bar{z}%
=0.$It therefore follows that after solving the first equation of system
(\ref{N1.13ab}), one can recursively sole the remaining two equations. In
particular, it is easy to see that the three vector elements%
\begin{equation}%
\begin{array}
[c]{c}%
\tilde{\psi}_{0}=(-v,u,-2\bar{z}_{x})^{\intercal},\text{ \ \ }\mathcal{\bar
{L}}_{0}=0;\\
\tilde{\psi}_{\theta}=(-u_{x}/\bar{z}_{x},1/\bar{z}_{x},(u_{x}^{2}%
-2v_{x})/(2\bar{z}_{x}^{2}))^{\intercal},\mathcal{\bar{L}}_{\theta}=0;\\
\tilde{\psi}_{\eta}=(u/2,-x/2,\partial^{-1}[(2v_{x}-u_{x}^{2})/(2\bar{z}%
_{x})]),\mathcal{\bar{L}}_{\eta}=(D_{x}\tilde{\psi}_{\eta},\bar{K}%
)-H_{\vartheta},
\end{array}
\label{N1.13bb}%
\end{equation}
are solutions of the system (\ref{N1.13ab}). The first two elements of
(\ref{N1.13bb}) lead to the Volterra symmetric vectors $\bar{\psi}_{0}%
=D_{x}\tilde{\psi}_{0},\bar{\psi}_{\theta}=D_{x}\tilde{\psi}_{\theta}\in
T^{\ast}(\bar{M}^{3}):$ $\bar{\psi}_{0}^{\prime}=\bar{\psi}_{0}^{\prime,\ast
},\bar{\psi}_{\theta}^{\prime}=\bar{\psi}_{\theta}^{\prime,\ast}$ entailing
the trivial conservation laws $(\bar{\psi}_{0},\bar{K})=0=(\bar{\psi}_{\theta
},\bar{K}).$ The third element of (\ref{N1.13bb}) gives rise to the Volterra
asymmetric vector $\bar{\psi}_{\eta}:=D_{x}\tilde{\psi}_{\eta}:\bar{\psi
}_{\eta}^{\prime}\neq\bar{\psi}_{\eta}^{\prime,\ast},$ entailing the following
inverse co-symplectic functional expression:%
\begin{equation}
\bar{\eta}^{-1}:=\bar{\psi}_{\eta}^{\prime}-\bar{\psi}_{\eta}^{\prime,\ast
}=\left(
\begin{array}
[c]{ccc}%
\partial & 0 & -\partial\frac{u_{x}}{\bar{z}_{x}}\\
0 & 0 & \partial\frac{1}{\bar{z}_{x}}\\
-\frac{u_{x}}{\bar{z}_{x}}\partial & \frac{1}{\bar{z}_{x}}\partial &
\begin{array}
[c]{c}%
\frac{u_{x}}{2\bar{z}_{x}}\partial\frac{u_{x}}{\bar{z}_{x}}-\\
-\frac{v_{x}}{\bar{z}_{x}}\partial\frac{1}{\bar{z}_{x}}-\frac{1}{\bar{z}_{x}%
}\partial\frac{v_{x}}{\bar{z}_{x}}%
\end{array}
\end{array}
\right)  . \label{N1.13ac}%
\end{equation}
Correspondingly, the Poissonian operator $\bar{\eta}:T^{\ast}(\bar{M}%
^{3})\rightarrow T(\bar{M}^{3})$ is
\begin{equation}
\bar{\eta}=\left(
\begin{array}
[c]{ccc}%
\partial^{-1} & u_{x}\partial^{-1} & 0\\
\partial^{-1}u_{x} & v_{x}\partial^{-1}+\partial^{-1}v_{x} & \partial^{-1}%
\bar{z}_{x}\\
0 & \bar{z}_{x}\partial^{-1} & 0
\end{array}
\right)  , \label{N1.13bc}%
\end{equation}
subject to which the following Hamiltonian representation
\begin{equation}
\bar{K}[u,v,\bar{z}]=-\bar{\eta}\text{ }\mathrm{grad}\text{ }H_{\eta
}|_{z=z_{x}^{2}} \label{N1.13cc}%
\end{equation}
holds on the manifold $\bar{M}^{3}$.

\subsection{Hamiltonian integrability analysis}

Next, we return to our integrability analysis of the dynamical system
(\ref{N1.3b}) on the functional manifold $M^{3}.$ It is easy to recalculate
the form of the Poissonian operator (\ref{N1.13bc}) on the manifold $\bar
{M}^{3}$ to that acting on the manifold $M^{3},$ giving rise to the second
Hamiltonian representation of (\ref{N1.3b}): \
\begin{equation}
K[u,v,z]=-\eta\text{ }\mathrm{grad}\text{ }H_{\vartheta}, \label{N1.13ca}%
\end{equation}
where $\eta:T^{\ast}(M^{3})\rightarrow T(M^{3})$ is the corresponding
Poissonian operator. As a next important \ point, the Poissonian operators
\ (\ref{N1.13}) and (\ref{N1.13bc}) are compatible \cite{FT,PM,BPS,Bl} on the
manifold $\bar{M}^{3}$; that is, the operator $\ $pencil $\ \ (\vartheta
+\lambda\eta):T^{\ast}(M^{3})\rightarrow T(M^{3})$ is also Poissonian for
arbitrary $\lambda\in\mathbb{R}.$ As a consequence, any operator of the form
\begin{equation}
\vartheta_{n}:=\vartheta(\vartheta^{-1}\eta)^{n} \label{N1.13bd}%
\end{equation}
for all $n\in\mathbb{Z}$ is Poissonian on the manifold $M^{3}$. Using now the
homotopy formula (\ref{N1.9}) and recursion property of the Poissonian pair
(\ref{N1.13a}) and (\ref{N1.13bc}), it is easy to construct the related
infinite hierarchy of mutually commuting conservation laws
\begin{equation}%
\begin{array}
[c]{c}%
\gamma_{j}=\int_{0}^{1}d\mu(\mathrm{grad}\text{ }\gamma_{j}[\mu u,\mu v,\mu
z],(u,v,z)^{\intercal}),\\
\mathrm{grad}\text{ }\gamma_{j}[u,v,z]:=\Lambda^{j}\mathrm{grad}\text{
}H_{\eta},
\end{array}
\label{N1.13d}%
\end{equation}
for the dynamical system (\ref{N1.3b}), where $j\in\mathbb{Z}_{+}$ and
$\ \Lambda:=\vartheta^{-1}\eta:T^{\ast}(M^{3})\rightarrow T^{\ast}(M^{3})$ is
the corresponding recursion operator, which satisfies the so called associated
Lax commutator relationship
\begin{equation}
d\Lambda/dt=[\Lambda,K^{\prime,\ast}]. \label{N1.13e}%
\end{equation}
In the course of above analysis and observations, we have proved the following result.

\begin{proposition}
\label{Pr_N1.1} The Riemann hydrodynamic system (\ref{N1.3b}) is a
bi-Hamiltonian dynamical system on the functional manifold $M^{3}$ with
respect to the compatible Poissonian structures $\vartheta,\eta:T^{\ast}%
(M^{3})$ $\rightarrow T(M^{3})$
\begin{equation}
\vartheta:=\left(
\begin{array}
[c]{ccc}%
0 & 1 & 0\\
-1 & 0 & 0\\
0 & 0 & 2z^{1/2}D_{x}z^{1/2}%
\end{array}
\right)  ,\eta:=\left(
\begin{array}
[c]{ccc}%
\partial^{-1} & u_{x}\partial^{-1} & 0\\
\partial^{-1}u_{x} & v_{x}\partial^{-1}+\partial^{-1}v_{x} & \partial
^{-1}z_{x}-2z\\
0 & z_{x}\partial^{-1}+2z & 0
\end{array}
\right)  \label{N1.13f}%
\end{equation}
and possesses an infinite hierarchy of mutually commuting conservation laws
(\ref{N1.13d}).
\end{proposition}

Concerning the existence of an additional infinite and parametrically
$\mathbb{R}\ni\lambda$-ordered hierarchy of conservation laws for the
dynamical system \ (\ref{N1.1a}), it is instructive to consider the dispersive
nonlinear dynamical system
\begin{equation}
\left(
\begin{array}
[c]{c}%
du/d\tau\\
dv/d\tau\\
dz/d\tau
\end{array}
\right)  =-\vartheta\text{ }\mathrm{grad}\text{ }H_{0}[u,v,z]:=\left(
\begin{array}
[c]{c}%
-(z^{-1/2})_{x}\\
-(u_{x}z^{-1/2})_{x}\\
z^{1/2}(\frac{u_{x}^{2}-2v_{x}}{2z})_{x}%
\end{array}
\right)  =\tilde{K}[u,v,z]. \label{N1.14}%
\end{equation}
By solving the corresponding Lax equation%
\begin{equation}
d\tilde{\varphi}/dt+\tilde{K}^{\prime,\ast}\tilde{\varphi}=0 \label{N1.15}%
\end{equation}
for an element $\ \tilde{\varphi}\in$ $T^{\ast}(M^{3})$ \ in a suitably chosen
asymptotic form, one can construct an infinite ordered hierarchy of
conservation laws for (\ref{N1.1a}), which we will not delve into here. This
hierarchy and the existence of an infinite and parametrically $\mathbb{R}%
\ni\lambda$-ordered hierarchy of conservation laws for the Riemann type
dynamical system \ (\ref{N1.1a}) provided compelling indications that it is
completely integrable in the sense of Lax on the functional manifold $M^{3}$.
We will be the complete integrability in the next section using rather
powerful differential-algebraic tools that were devised recently in
\cite{PAPP,PoP,PAPP-1}.

\bigskip

\section{Differential-algebraic integrability analysis: $N=3$}

Consider a polynomial differential ring $\mathcal{K}\{u\}\subset
\mathcal{K}:=\mathbb{R}\{\{x,t\}\}$ generated by a fixed functional variable
$u\in\mathbb{R}\{\{x,t\}\}$ and invariant with respect to two differentiations
$D_{x}:=\partial/\partial x$ and $D_{t}:=\partial/\partial t+u\partial
/\partial x$ that satisfy the Lie-algebraic commutator relationship
(\ref{N1.1a})%

\begin{equation}
\lbrack D_{x},D_{t}]=u_{x}D_{x} \label{N2.1}%
\end{equation}
together with the constraint (\ref{N1.3a}) expressed in the
differential-algebraic functional form
\[
D_{t}^{3}u=-2D_{t}^{2}uD_{x}u.
\]

Since the Lax representation for the dynamical system \ (\ref{N1.3b}) can be
interpreted \cite{PAPP,BPS} as the existence of a finite-dimensional invariant
ideal $\mathcal{I}\{u\}\subset\mathcal{K}\{u\}$ realizing the corresponding
finite-dimensional representation of the the Lie-algebraic commutator
relationship \ (\ref{N2.1}), this ideal can be constructed as
\begin{equation}
\mathcal{I}\{u\}:=\{\lambda^{2}uf_{1}+\lambda vf_{2}+z^{1/2}f_{3}%
\in\mathcal{K}\{u\}:f_{j}\in\mathcal{K},1\leq j\leq3,\lambda\in\mathbb{R\}}%
,\label{N2.2}%
\end{equation}
where $v=D_{t}u,z=D_{t}^{2}u$ and $\lambda\in\mathbb{R}$ is an arbitrary real
parameter. To find finite-dimensional representations of the $D_{x}$- and
$D_{t}$-differentiations, it is necessary \cite{PAPP} first to find the
$D_{t}$-invariant kernel $\ker D_{t}\subset\mathcal{I}\{u\}$ and next to check
its invariance with respect to the $D_{x}$-differentiation. It is easy to show
that
\begin{equation}
\ker D_{t}=\{f\in\mathcal{K}^{3}\{u\}:D_{t}f=q(\lambda)f,\text{ \ \ }%
\lambda\in\mathbb{R}\},\label{N2.3}%
\end{equation}
where the matrix $q(\lambda):=q[u,v,z;\lambda]\in$\textrm{End} $\mathcal{K}%
\{u\}^{3}$ is given as
\begin{equation}
q(\lambda)=\left(
\begin{array}
[c]{ccc}%
0 & 0 & 0\\
-\lambda & 0 & 0\\
0 & -\lambda z_{x} & u_{x}%
\end{array}
\right)  .\label{N2.4}%
\end{equation}

To obtain the corresponding representation of the $D_{x}$-differentiation in
the space $\mathcal{K}^{3},$ it suffices to find a matrix $l(\lambda
):=l[u,v,z;\lambda]\in$\textrm{End} $\mathcal{K}\{u\}^{3}$ that
\begin{equation}
D_{x}f=l(\lambda)f\label{N2.5}%
\end{equation}
for $f\in\mathcal{K}\{u\}^{3}$ and the related ideal
\begin{equation}
\mathcal{R}\{u\}:=\{<g,f>_{\mathcal{K}^{3}}:f\in\ker D_{t}\subset
\mathcal{K}^{3}\{u\},\ g\in\mathcal{K}^{3}\}\label{N2.6}%
\end{equation}
is $D_{x}$-invariant with respect to the matrfix differentiation
representation \ref{N2.5}). \ Straightforward calculations using this
invariance condition then yield the following matrix
\begin{equation}
l(\lambda)=\left(
\begin{array}
[c]{ccc}%
\lambda^{2}u\sqrt{z} & \lambda v\sqrt{z} & z\\
-\lambda^{3}tu\sqrt{z} & -\lambda^{2}tv\sqrt{z} & -\lambda tz\\%
\begin{array}
[c]{c}%
\lambda^{4}(tuv-u^{2})-\\
-\lambda^{2}u_{x}/\sqrt{z}%
\end{array}
&
\begin{array}
[c]{c}%
-\lambda v_{x}/\sqrt{z}+\\
+\lambda^{3}(tv^{2}-uv)
\end{array}
&
\begin{array}
[c]{c}%
\lambda^{2}\sqrt{z}(u-tv)-\\
-z_{x}/2z
\end{array}
\end{array}
\right)  \label{N2.7}%
\end{equation}
entering the linear equation \ref{N2.5}).  Thus, the following proposition  is proved.

\begin{proposition}
The generalized Riemann type dynamical system \ (\ref{N1.3a}) is a
bi-Hamiltonian integrable flow \ possessing  a  non-autonomous Lax
representation of the form
\begin{align}
D_{t}f &  =\left(
\begin{array}
[c]{ccc}%
0 & 0 & 0\\
-\lambda & 0 & 0\\
0 & -\lambda z_{x} & u_{x}%
\end{array}
\right)  f,\text{ }\label{N2.7a}\\
D_{x}f &  =\left(
\begin{array}
[c]{ccc}%
\lambda^{2}u\sqrt{z} & \lambda v\sqrt{z} & z\\
-\lambda^{3}tu\sqrt{z} & -\lambda^{2}tv\sqrt{z} & -\lambda tz\\%
\begin{array}
[c]{c}%
\lambda^{4}(tuv-u^{2})-\\
-\lambda^{2}u_{x}/\sqrt{z}%
\end{array}
&
\begin{array}
[c]{c}%
-\lambda v_{x}/\sqrt{z}+\\
+\lambda^{3}(tv^{2}-uv)
\end{array}
&
\begin{array}
[c]{c}%
\lambda^{2}\sqrt{z}(u-tv)-\\
-z_{x}/2z
\end{array}
\end{array}
\right)  f,\nonumber
\end{align}
where $\lambda\in\mathbb{R}$ is an arbitrary spectral parameter and $f\in
C^{(\infty)}(\mathbb{R}^{2};\mathbb{R}^{3}).$
\end{proposition}

\begin{remark}
Simple analogs of the above differential-algebraic calculations for the case
$N=2$ lead readily to the corresponding Riemann type hydrodynamic system
\begin{equation}
D_{t}u=\bar{z}_{x}^{2},\text{ \ \ \ }D_{t}\bar{z}=0\label{N2.7aa}%
\end{equation}
\ on the functional manifold $\bar{M}^{2}$, which possesses the following
matrix Lax representation:%
\begin{equation}
D_{t}f=\left(
\begin{array}
[c]{cc}%
0 & 0\\
-\lambda\bar{z}_{x} & u_{x}%
\end{array}
\right)  ,\text{ \ \ \ \ }D_{x}f=\left(
\begin{array}
[c]{cc}%
\bar{z}_{x} & 0\\
-\lambda(u+u_{x}/\bar{z}_{x} & -\bar{z}_{xx}/\bar{z}_{x}%
\end{array}
\right)  f,\label{N2.7ab}%
\end{equation}
where $\lambda\in\mathbb{R}$ is an arbitrary spectral parameter and $f\in
C^{(\infty)}(\mathbb{R}^{2};\mathbb{R}^{2}).$
\end{remark}

As one can readily see, these differential-algebraic results provide a direct
proof of Proposition \ref{Pr_N1.0} describing the integrability of system
(\ref{N1.3b}) for $N=3.$ The matrices (\ref{N2.7}) are not of standard form
since they depend explicitly on the temporal evolution parameter
$t\in\mathbb{R}.$ Nonetheless, the matrices (\ref{N2.4}) and (\ref{N2.7})
satisfy for all $\lambda\in\mathbb{R}$ the well-known Zakharov-Shabat type
compatibility condition
\begin{equation}
D_{t}l(\lambda)=[q(\lambda),l(\lambda)]+D_{x}l(\lambda)-u_{x}l(\lambda),
\label{N2.8}%
\end{equation}
which follows from the Lax type\ relationships (\ref{N2.3}) and (\ref{N2.5})
\begin{equation}
D_{t}f=q(\lambda)f,\text{ \ \ \ \ }D_{x}f=l(\lambda)f \label{N2.9}%
\end{equation}
and the commutator condition (\ref{N2.1}). Moreover, taking into account that
the dynamical system (\ref{N1.3b}) has a compatible Poissonian pair
(\ref{N1.13}) and (\ref{N1.13bc}) depending only on the variables
$(u,v,z)^{\intercal}\in M^{3}$ \ and not depending \ on the temporal variable
$t\in\mathbb{R},$ one can certainly assume that it also possesses a standard
autonomous Lax representation, which can possibly be found by means of a
suitable gauge transformation of (\ref{N2.9}). We plan to pursue this line of
analysis in a forthcoming paper.

\section{Concluding remarks}

A new nonlinear Hamiltonian dynamical system representing Riemann type
hydrodynamic equation (\ref{N1.1}) in two and three dimensions proves to be a
very interesting example of a Lax integrable dynamical system, as we have
proved here. In particular, the integrability prerequisites of this dynamical
system, such as compatible Poissonian structures, an infinite hierarchy of
conservation laws and related Lax representation have been constructed by
means of both the symplectic gradient-holonomic approach \cite{PM,BPS,HPP} and
innovative differential-algebraic tools devised recently \cite{PAPP,GBPPP} for
analyzing the integrability of a special infinite hierarchy of Riemann type
hydrodynamic systems. It is also quite clear from recent research in this area
and our work in this paper that the dynamical system (\ref{N1.1}) is a Lax
integrable bi-Hamiltonian flow for arbitrary integers $N\in\mathbb{N};$ this
is perhaps most readily verified by means of the differential-algebraic
approach, which was devised and successfully applied here for the cases $N=2$
and $3.$

We have seen in the course of this investigation that perhaps the most
important lesson that one can derive from this approach is the following: If
an investigation of a given nonlinear Hamiltonian dynamical system via the
gradient-holonomic method indicates (but does not necessarily prove) that the
system is Lax integrable, then its Lax representation, can often be shown to
exist and then successfully derived by means of a suitably constructed
invariant differential ideal $\mathcal{I}\{u\}$ of the ring $\mathcal{K}\{u\}$
in accordance with the differential-algebraic approach developed here for the
integrability analysis of the Riemann hydrodynamical system investigated
above. Consequently, when it comes applying this lesson to the investigation
of other nonlinear dynamical systems, it is natural to start with systems that
are known to be Lax integrable and to try to identify and characterize those
algebraic structures responsible for the existence of a related
finite-dimensional matrix representation for the basic $D_{x}$- and $D_{t}%
$-differentiations in a vector space $\mathcal{K}^{p}$ for some finite
$p\in\mathbb{Z}_{+}.$

It seems plausible that if one could do this for several classes of Lax
integrable dynamical systems, certain patterns in the algebraic structures may
be detected that can be used to assemble a more extensive array of symplectic
and differential-algebraic tools capable of resolving the question of complete
integrability for many other types of nonlinear Hamiltonian dynamical systems.
Moreover, if the integrability is established in this manner, the approach
should also serve as a means of constructing associated artifacts of the
integrability such as Lax representations and hierarchies of mutually
commuting invariants. As a particular differential-algebraic problem of
interest concerning these matrix representations, one can seek to develop a
scheme for the effective construction of functional generators of the
corresponding invariant finite-dimensional ideals $\mathcal{I}\{u\}\subset
\mathcal{K}\{u\}$ under given differential-algebraic constraints imposed on
the $D_{x}$- and $D_{t}$-differentiations.

We have demonstrated here that an approach combining the gradient-holonomic
method with some recently devised differential-algebraic techniques can be a
very effective and efficient way of investigating integrability for a
particular class of infinite-dimensional Hamiltonian dynamical systems
(generalized Riemann hydrodynamical systems). But a closer look at the
specific details of the approach employed here reveals, we believe, that this
combination of methods can be adapted to perform effective integrability
analysis of a much wider range of dynamical systems - a goal that we intend to
pursue in the very near future.

\section*{Acknowledgements}

The authors are much obliged to Prof. M.Pavlov (P.N. Lebedev Physical
Institute of RAS and M. Lomonosov State University, Moscow, Russian Federation
) for very instrumental discussion of the work, valuable advice, comments and
remarks. They are also grateful to Prof. Z. Popowicz (Wroc\l aw University,
Poland) for fruitful cooperation. Special thanks are due the Scientific and
Technological Research Council of Turkey (TUBITAK-2011) for partial support of
the research by A.K. Prykarpatsky and Y.A. Prykarpatsky, and the National
Science Foundation (Grant CMMI-1029809) for partial support of the research of
D. Blackmore.

\end{document}